\documentclass[aps,pra,twocolumn,superscriptaddress,longbibliography]{revtex4-1}

\usepackage{amsmath}
\usepackage{amsfonts}
\usepackage{amssymb}
\usepackage{graphicx}
\usepackage{color}
\usepackage[colorlinks=true, citecolor=blue]{hyperref}

\newcommand{\Tr}{\mbox{Tr}}

\newcommand{\ket}[1]{\left|#1\right\rangle}

\newcommand{\qed}{\nobreak \ifvmode \relax \else
      \ifdim\lastskip<1.5em \hskip-\lastskip
      \hskip1.5em plus0em minus0.5em \fi \nobreak
      \vrule height0.75em width0.5em depth0.25em\fi}

\begin{document}
\title{Robust quantum sensing with strongly interacting probe systems}

\author{Shane Dooley}
\email[]{dooleysh@gmail.com}
\affiliation{National Institute of Informatics, 2-1-2 Hitotsubashi, Chiyoda-ku, Tokyo 101-8430, Japan.}

\author{Michael Hanks}
\affiliation{Department of Informatics, School of Multidisciplinary Sciences, Sokendai (The Graduate University for Advanced Studies), 2-1-2 Hitotsubashi, Chiyoda-ku, Tokyo 101-8430 Japan}
\affiliation{National Institute of Informatics, 2-1-2 Hitotsubashi, Chiyoda-ku, Tokyo 101-8430, Japan.}

\author{Shojun Nakayama}
\affiliation{National Institute of Informatics, 2-1-2 Hitotsubashi, Chiyoda-ku, Tokyo 101-8430, Japan.}

\author{William J. Munro}
\affiliation{NTT Research Center for Theoretical Quantum Physics, NTT Corporation, 3-1 Morinosato-Wakamiya, Atsugi 243-0198, Japan.}
\affiliation{NTT Basic Research Laboratories, 3-1 Morinosato-Wakamiya, Atsugi, Kanagawa 243-0198, Japan.}
\affiliation{National Institute of Informatics, 2-1-2 Hitotsubashi, Chiyoda-ku, Tokyo 101-8430, Japan.}

\author{Kae Nemoto}
\affiliation{National Institute of Informatics, 2-1-2 Hitotsubashi, Chiyoda-ku, Tokyo 101-8430, Japan.}
\affiliation{Department of Informatics, School of Multidisciplinary Sciences, Sokendai (The Graduate University for Advanced Studies), 2-1-2 Hitotsubashi, Chiyoda-ku, Tokyo 101-8430 Japan}

\date{\today}

\begin{abstract}
In the field of quantum metrology and sensing, a collection of quantum systems (e.g. spins) are used as a probe to estimate some physical parameter (e.g. magnetic field). It is usually assumed that there are no interactions between the probe systems. We show that strong interactions between them can increase robustness against thermal noise, leading to enhanced sensitivity. In principle, the sensitivity can scale exponentially in the number of probes -- even at non-zero temperatures -- if there are long-range interactions. This scheme can also be combined with other techniques, such as dynamical decoupling, to give enhanced sensitivity in realistic experiments.
\end{abstract}


\maketitle

\section{Introduction}

The estimation of physical quantities or parameters is a crucial task in science. The field of quantum metrology and sensing aims to exploit quantum coherence or entanglement to give highly sensitive estimates of such quantities \cite{Gio-06, Deg-17}. Known applications include time and frequency estimation \cite{Lud-15}, gravitational wave detection \cite{Cav-81, Aas-13}, magnetometry \cite{Bud-07, Ron-14, Tan-15} and electrometry \cite{Dol-11}. In a typical quantum sensing scheme, $N$ probe systems evolve for a sensing time $t$, picking up a dependence on the physical parameter of interest, before readout. This procedure is repeated $\nu = T/t$ times during a total available measurement time $T$, and an estimate of the parameter is inferred from the accumulated measurement data. However, the quantum coherence of the probe decays on a timescale denoted $T_2$. This limits the useful sensing time $t \lesssim T_2$, which in turn limits the sensitivity of the final estimate. In principle, dynamical decoupling \cite{Vio-99} or other techniques \cite{Ave-16, Mat-18} can be used to extend the coherence time to its fundamental limit $T_2 \leq 2 T_1$, where $T_1$ is the probe relaxation time. It thus appears that the sensitivity is limited by the probe relaxation time $T_1$. However, it is usually assumed that the $N$ probe systems are not interacting. In this paper we show that the $T_1$ sensitivity limit with non-interacting probes can be overcome with interacting probes. Our scheme is based on the idea that strong interactions can modify the energy level structure of a quantum system so that dissipation tends to drive the system into a multidimensional ground space where quantum information can be stored robustly despite energy relaxation \cite{Paz-98, Ahn-02, Sar-05, Ipp-15,Pat-11,Bro-16,Rei-17,Leg-13,Coh-14,Bar-00, Kap-16}. 

\begin{figure*}
\includegraphics[width=2\columnwidth]{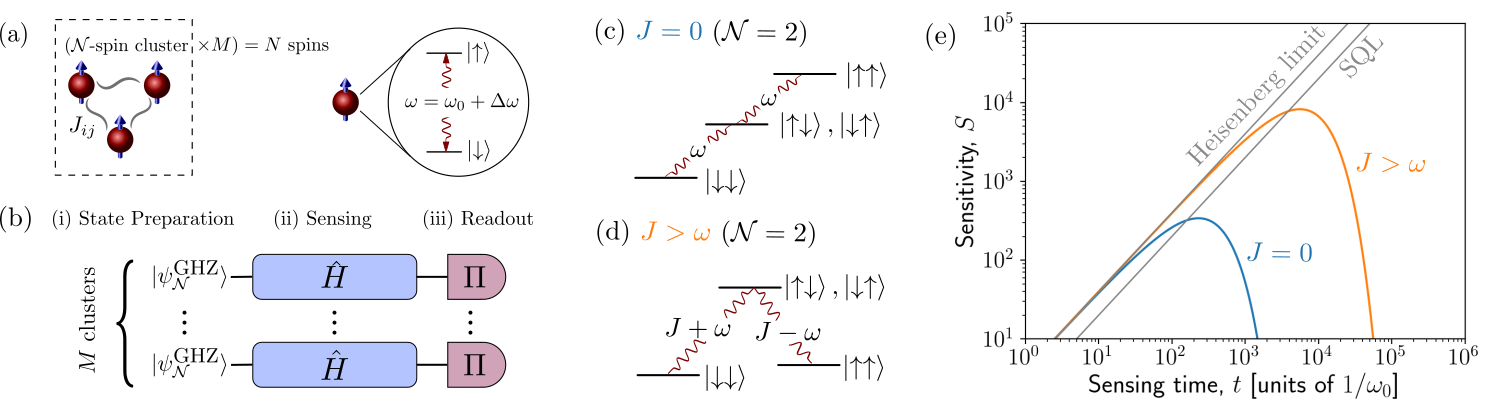} 
\caption{(a) We consider estimation of $\omega$ -- the frequency gap between spin states, $\ket{\uparrow}$ and $\ket{\downarrow}$ -- in a system of $N$ spins (divided into $M$ clusters of $\mathcal{N}$ spins). (c) In the simplest case ($\mathcal{N}=2$), non-interacting spins ($J=0$) that are coupled to a thermal environment will, at zero-temperature, decay to the ground state $\ket{\downarrow\downarrow}$, losing all information about the parameter $\omega$. (d) A strong ferromagnetic Ising interaction ($J>\omega$) will modify the energy level structure so that, at zero-temperature, information about $\omega$ can be encoded in the relative phase of the two ground states $\ket{\uparrow\uparrow}$ and $\ket{\downarrow\downarrow}$. (e) Comparing the orange line to the blue line shows that, even for non-zero temperatures, the sensitivity is enhanced when the spins are strongly interacting [plotted for coupling strength $J = 5\omega_0$, inverse temperature $\beta = 1 / \hbar\omega_0$, and Ohmic spectral density $f(\Omega) = 0.001 \times \Omega$.] \label{fig:intro}}
\end{figure*}

We focus on the problem of estimating the resonant frequency $\omega$ between two spin-$1/2$ states $\ket{\uparrow}$ and $\ket{\downarrow}$, given a probe consisting of $N$ spin-$1/2$ particles. If there is no decoherence the sensitivity usually scales as $S \propto t$, where $S = 1/T(\delta\omega)^2$ and $\delta\omega$ is the error of the frequency estimate \cite{Deg-17}. For example, if we are restricted to the preparation of separable spin states, the optimal sensitivity (known as the standard quantum limit) is $S_\text{SQL} = N t$. If entangled states are allowed the sensitivity can, in principle, be increased to the Heisenberg limit $S_\text{HL} = N^2 t$, a factor of $N$ enhancement compared to the standard quantum limit. In practice, however, even if dynamical decoupling is employed, energy relaxation will prevent the sensitivity from increasing indefinitely with increasing sensing time $t$. This means that the sensitivity $S(t)$ can -- at best -- approach the Heisenberg limit only for relatively short times $t$ and will eventually reach a maximum value $\max_t S (t)$ at some optimal time $t_\text{opt}$, before decreasing as the spins thermalize [for example, see Fig. \ref{fig:intro}]. However, a strong ferromagnetic interaction between the spins can lead to an increased $t_\text{opt}$ and thus an enhanced estimate of $\omega$. The simplest example of this idea is illustrated in Fig. \ref{fig:intro}(c, d, e) for $N=2$ interacting spins.

We structure the paper as follows. We begin the Results section by describing our model and our frequency estimation scheme. We then derive the sensitivity corresponding to the estimation scheme and show how it varies depending on the strength of interactions among the probe spins. We will see that for strong ferromagnetic interactions the sensitivity increases exponentially with decreasing temperature. With long-range ferromagnetic interactions between the spins it is also possible, in principle, to achieve a sensitivity that scales exponentially in the number of probe spins $N$, even at non-zero environment temperatures. We conclude with a discussion of our results.

\section{Results}

\subsection{Model}

Our measurement probe consists of $N$ spin-1/2 particles. We divide the $N$ particles into $M$ identical clusters of size $\mathcal{N} = N / M$ and we perform identical, independent experiments in parallel on each $\mathcal{N}$-spin cluster. Each cluster evolves by the Hamiltonian $\hat{H} = \hat{H}_\text{spins} + \hat{H}_\text{env} + \hat{H}_\text{int}$, where: \begin{eqnarray} \hat{H}_\text{spins} &=& \frac{\hbar\omega}{2} \sum_{i=1}^{\mathcal{N}} \hat{\sigma}^z_{i} - \frac{\hbar}{4} \sum_{i,j} J_{i,j} \hat{\sigma}^z_{i} \otimes \hat{\sigma}^z_{j} , \label{eq:H_spins} \\ \hat{H}_\text{env} &=& \hbar \sum_{i=1}^{\mathcal{N}} \sum_k \Omega_k \hat{a}_{i,k}^\dagger \hat{a}_{i,k} , \label{eq:H_env} \\ \hat{H}_\text{int} &=& \hbar \sum_{i=1}^{\mathcal{N}} \hat{\sigma}^x_{i} \otimes \hat{E}_i . \label{eq:H_int} \end{eqnarray} Here $\omega = \omega_{0} + \Delta\omega$ and we would like to estimate $\Delta\omega$, a small unknown deviation from the known frequency $\omega_{0}$. The strength of the Ising interaction between the $i$'th and $j$'th spins in each cluster is $J_{i,j}$. To model energy relaxation, each spin has a dipole-dipole coupling to an environment of harmonic oscillators (indexed by $k$) via the environment operator $\hat{E}_i \equiv \sum_k \lambda_k (\hat{a}_{i,k}^\dagger + \hat{a}_{i,k})$. We assume that the environment is in a thermal state $\hat{\rho}_\text{env} \propto e^{-\beta \hat{H}_\text{env}}$ with inverse temperature $\beta = 1/k_B T_\text{env}$, where $k_B$ is the Boltzmann constant and $T_\text{env}$ is the environment temperature. 
\subsection{Frequency estimation scheme} 

We divide our frequency estimation scheme into the following four stages [see Fig. \ref{fig:intro}(b)]:

(i) State Preparation. The $\mathcal{N}$-spin cluster is prepared in the entangled Greenberger-Horne-Zeilinger (GHZ) state: \begin{equation} |\psi^\text{GHZ}_{\mathcal{N}} \rangle = \frac{1}{\sqrt{2}} \left(\ket{\uparrow}^{\otimes \mathcal{N}} + \ket{\downarrow}^{\otimes \mathcal{N}} \right) . \label{eq:GHZ_mu} \end{equation} 

(ii) Sensing. The cluster evolves by the Hamiltonian $\hat{H}$, picking up a dependence on the unknown parameter $\omega$. The reduced state of the cluster after a sensing time $t$ is $\hat{\rho} (t)$.

(iii) Readout. The $\mathcal{N}$-spin cluster is measured with the POVM $\Pi = \{ \hat{\Pi}_0 , \hat{\Pi}_1 \}$, where: \begin{equation} \hat{\Pi}_{0} = \frac{1}{2} + \frac{1}{2} \left( \hat{\Lambda} e^{-i\phi} + \hat{\Lambda}^{\dagger} e^{i\phi} \right) , \quad \hat{\Pi}_{1} = \hat{\mathbb{I}} - \hat{\Pi}_{0} . \label{eq:easier_POVM} \end{equation} Here $\hat{\Lambda} = (\hat{\sigma}^{-})^{\otimes \mathcal{N}}$ and $\phi$ is a controllable parameter that determines the measurement bias point \cite{Deg-17}. This POVM corresponds to a binary measurement in the subspace spanned by the states $\ket{\uparrow}^{\otimes \mathcal{N}}$ and $\ket{\downarrow}^{\otimes \mathcal{N}}$ that make up the initial GHZ state $\ket{\psi_{\mathcal{N}}^\text{GHZ}}$. The measurement leads to the outcome ``0'' with probability $p = \Tr [ \hat{\rho}(t) \hat{\Pi}_0]$ or the outcome ``1'' with probability $1-p$.

(iv) Repetition. Steps (i)--(iii) are repeated on each cluster for a total time $T$ giving $\nu = T / t$ repetitions.

We define the sensitivity as $S = 1 / T(\delta\omega)^2$, where $\delta\omega$ is the root-mean-squared error of the frequency estimate. The Cramer-Rao inequality $(\delta\omega)^2 \geq 1/ (M\nu F)$ gives an upper bound for the error of the frequency estimate \cite{Gio-06, Deg-17}, where \begin{equation} F = \frac{\left| \partial p / \partial \omega \right|^2}{p (1 - p)} , \label{eq:binary_FI} \end{equation} is the (classical) Fisher information corresponding to the binary measurement of the $\mathcal{N}$-spin cluster. In the limit of many repetitions $\nu \gg 1$ it is possible to saturate the Cramer-Rao bound with maximum likelihood estimation \cite{Bra-92}. Substituting $\nu = T/t$ we thus obtain the formula $S = MF/t$ for the sensitivity. In the next section we calculate the Fisher information $F$, and hence the sensitivity $S$ for the frequency estimation scheme described above.

\subsection{Calculating the sensitivity}

From the Hamiltonian given in Eqs. \ref{eq:H_spins}--\ref{eq:H_int}, a standard derivation \cite{Sch-07} leads to the Born-Markov master equation for the reduced state of the $\mathcal{N}$-spin cluster (see the Supplementary Information for details): \begin{eqnarray} \frac{d}{dt}\hat{\rho}(t) = -\frac{i}{\hbar}\Big[ \hat{H}_\text{spins} , && \hat{\rho}(t) \Big] \nonumber\\  + \int_0^\infty d\tau \sum_{i=1}^{\mathcal{N}} \Big\{ \mathcal{C}(\tau) &\Big[& \hat{\sigma}_i^x(-\tau)\hat{\rho}(t) ,\hat{\sigma}_i^x(0) \Big] \nonumber\\ \qquad\qquad + \mathcal{C}(-\tau) &\Big[& \hat{\sigma}_i^x(0) , \hat{\rho}(t) \hat{\sigma}_i^x(-\tau) \Big] \Big\} , \label{eq:BM_ME} \end{eqnarray} where $\mathcal{C}(\tau) \equiv \Tr\{ \hat{E}_i(\tau)\hat{E}_i(0) \hat{\rho}_\text{env} \}$ is the environment self-correlation function and $\hat{\sigma}_i^x(\tau) \equiv e^{i\tau\hat{H}_\text{spins}/\hbar} \hat{\sigma}_i^x e^{-i\tau\hat{H}_\text{spins}/\hbar}$, $\hat{E}_i(\tau) \equiv e^{i\tau\hat{H}_\text{env}/\hbar} \hat{E} e^{-i\tau\hat{H}_\text{env}/\hbar}$.

Taking the expectation value of the master Eq. \ref{eq:BM_ME} with the operator $\hat{\Lambda}$ gives (after a rotating wave approximation -- see the Supplementary Information for details) the equation of motion: \begin{equation} \frac{d}{dt} \langle \hat{\Lambda}\rangle = \mathcal{N} \left( - i \omega - \Gamma / 2 \right) \langle \hat{\Lambda} \rangle . \label{eq:E_o_M} \end{equation} Here the average decay rate is $\Gamma = (1/\mathcal{N}) \sum_{i=1}^{\mathcal{N}} \xi_i$, where \begin{equation} \xi_i = 2 \, \text{Re} \int_0^\infty d\tau \Big[ \mathcal{C}(\tau) e^{-i\tau (\mathcal{J}_i - \omega) } + \mathcal{C}(-\tau) e^{i\tau (\mathcal{J}_i + \omega) } \Big] , \label{eq:xi_i} \end{equation} is the decay rate associated with the $i$'th spin. We ignore the imaginary part of the integral in Eq. \ref{eq:xi_i}, since it leads to a negligible frequency shift. In the equation for $\xi_i$ above we have introduced $\mathcal{J}_i \equiv \sum_{j=1, j \neq i}^{\mathcal{N}} J_{i,j}$, which is the collective coupling strength of the $i$'th spin to all other spins in the $\mathcal{N}$-spin cluster. We will see below that the size of this collective coupling strength $\mathcal{J}_i$ relative to the spin frequency $\omega$ is a key parameter in determining the relaxation dynamics of the spin system. 

The equation of motion Eq. \ref{eq:E_o_M} is easily solved for $\langle  \hat{\Lambda}(t) \rangle$ and the solution is substituted into $p = \Tr [\hat{\rho} (t) \hat{\Pi}_0]$ to calculate the probability $p$. For the initial state given in Eq. \ref{eq:GHZ_mu} we find that: \begin{equation} p = \frac{1}{2} + \frac{1}{2}\cos\left( \omega \mathcal{N} t + \phi \right) e^{- \mathcal{N} \Gamma t / 2 }  . \end{equation} Now, we can find an expression for the classical Fisher information $F$ by substituting our solution for $p$ into Eq. \ref{eq:binary_FI}. Choosing the measurement bias point $\phi = \frac{\pi}{2} - \mathcal{N} \omega_0 t$ gives $F = \mathcal{N}^2 t^2 e^{- \mathcal{N} \Gamma t }$ so that the total sensitivity of the frequency estimate is: \begin{equation} S = MF/t = N \mathcal{N} t e^{- \mathcal{N} \Gamma t } .  \end{equation} If $\Gamma \neq 0$, we can optimise over $t$ to obtain: \begin{equation} \max_{ t } S  = \frac{N}{e\Gamma} , \quad t_\text{opt} = \frac{1}{\mathcal{N}\Gamma} , \label{eq:max_sens} \end{equation} where $e \approx 2.7$ is the Euler number and the optimum occurs at the time $t_\text{opt}$. 

\subsection{Calculating the average decay rate}

It is clear that the sensitivity depends crucially on the average decay rate $\Gamma = (1/\mathcal{N})\sum_{i=1}^{\mathcal{N}}\xi_i$, which in turn depends on the individual decay rates $\xi_i$. We can calculate $\xi_i$ by computing the integrals in Eq. \ref{eq:xi_i}. The result depends on the strength of the collective coupling $\mathcal{J}_i$ relative to the spin frequency $\omega$. Assuming $\omega > 0$,  we find the following three possibilities (see the Supplementary Information for details):

(i) If $-\omega < \mathcal{J}_i < \omega$ (weak coupling) we have: \begin{equation} \xi_i = \gamma_i^- (\bar{n}_i^- + 1) + \gamma_i^+ \bar{n}_i^{+} . \label{eq:WC} \end{equation} 

(ii) If $\mathcal{J}_i < -\omega$ (strong anti-ferromagnetic coupling): \begin{equation} \xi_i = \gamma_i^{-} (\bar{n}_i^{-} + 1) + \gamma_i^{+} (\bar{n}_i^{+} + 1) . \end{equation} 

(iii) If $\mathcal{J}_i > \omega$ (strong ferromagnetic coupling): \begin{equation} \xi_i = \gamma_i^{-} \bar{n}_i^{-} + \gamma_i^{+} \bar{n}_i^{+} . \label{eq:SFC} \end{equation} 

\noindent Here $\bar{n}_i^{\pm} = 1 / (e^{\hbar\beta |\mathcal{J}_i \pm \omega|} - 1)$ is the thermal occupation of the environment oscillator with frequency $|\mathcal{J}_i \pm \omega|$, and we have defined $\gamma_i^{\pm} = 2\pi f(|\mathcal{J}_i \pm \omega|)$ where $f(\Omega)$ is the environment spectral density.

We can immediately see that the strong ferromagnetic coupling regime is of particular interest, since at zero-temperature ($\beta\to\infty \implies \bar{n}_i^{\pm} \to 0$) the decay rate $\xi_i$ vanishes for strong ferromagnetic coupling (but is non-zero for weak coupling or for strong anti-ferromagnetic coupling). This zero-temperature behaviour is an indication that at low, but non-zero temperatures there is a qualitative difference between the strong ferromagnetic case and the weak coupling or strong anti-ferromagnetic coupling. We now consider the implications of this for the sensitivity of our frequency estimation scheme, focussing on the example of a one-dimensional spin chain. 

\subsection{Example: a 1-d spin chain} \label{sec:1-d}

The analysis so far has been very general (we have not specified the coupling strengths $J_{i,j}$). However, to gain further insight we focus on a concrete example: a one-dimensional spin chain with the interaction $J_{i,j} = J |i - j|^{-\alpha}$, where $|i-j|$ is the distance between the $i$'th and $j$'th spin. Here $|i-j|$ takes values from the set $\{1,2,...,\mathcal{N}\}$ and $\alpha$ controls the range of the interaction; small $\alpha$ corresponding to long-range interaction and large $\alpha$ to short-range interaction. We choose this form for $J_{i,j}$ because it covers a broad range of interesting examples including the infinite range interaction ($\alpha = 0$; also known as one-axis twisting), Coulomb-like interaction ($\alpha=1$), dipole-dipole interaction ($\alpha = 3$), nearest-neighbour interaction ($\alpha\to\infty$), and also the case of no interaction ($J = 0$). Moreover, it can be implemented experimentally for $0 \leq \alpha \leq 3$ with trapped ions \cite{Bri-12, Isl-13, Ric-14, Jur-14}. A necessary criterion for enhanced sensitivity in our scheme is that, for each spin, the collective coupling should be larger than spin frequency, $\mathcal{J}_i > \omega$ for all $i$ (see Eq. \ref{eq:SFC}). To simplify the analysis, we assume that the spin chain has periodic boundary conditions. This is convenient because it results in a collective coupling $\mathcal{J} \equiv \mathcal{J}_i = \sum_{j=1,j\neq\lfloor\mathcal{N}/2\rfloor}^{\mathcal{N}} J |\lfloor\mathcal{N}/2\rfloor - j|^{-\alpha}$ that is independent of the spin label $i$, so that the condition $\mathcal{J} > \omega$ for strong ferromagnetic coupling is the same for each spin. (We note, however, that for open boundary conditions the results will be qualitatively similar provided that $\mathcal{J}_i > \omega$ for all $i$.) Since the collective coupling is the same for each spin we have that $\gamma_i^{\pm} = \gamma^{\pm}$ and $\bar{n}_i^{\pm} = \bar{n}^{\pm}$ are also independent of $i$. This means that the average decay rate is written simply as:

(i) For weak coupling: \begin{equation} \Gamma =  \gamma^- (\bar{n}^- + 1) + \gamma^+ \bar{n}^+ , \end{equation}

(ii) For strong anti-ferromagnetic coupling: \begin{equation} \Gamma = \gamma^- (\bar{n}^- + 1) + \gamma^+ (\bar{n}^+ + 1) . \end{equation}

(iii) For strong ferromagnetic coupling: \begin{equation} \Gamma =  \gamma^- \bar{n}^- + \gamma^+ \bar{n}^+ . \end{equation}

\noindent Substituting into Eq. \ref{eq:max_sens}, gives simple expressions for the sensitivity in each case.

The only variables that affect the average decay rates are the inverse temperature $\beta$ (via the thermal occupation $\bar{n}^{\pm}$), the strength of the collective coupling $\mathcal{J}$ relative to $\omega$ (which enters through both $\bar{n}^{\pm}$ and $\gamma^{\pm}$), and the form of the spectral density $f(\Omega)$ (via the dissipation rate $\gamma^{\pm}$). We now examine the dependence of the sensitivity on these variables.

\begin{figure}
\begin{center}
\includegraphics[width=\columnwidth]{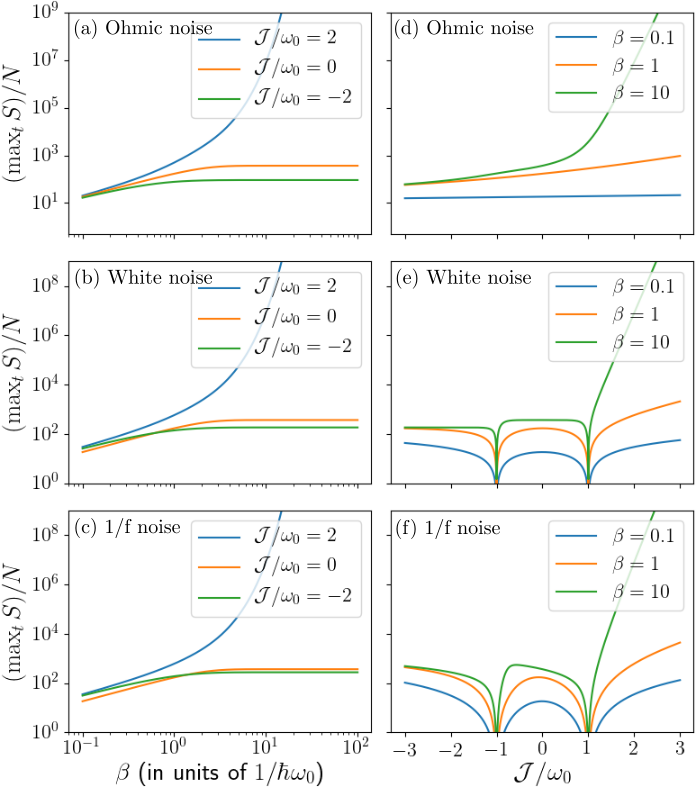} 
\caption{(a), (b), and (c) show that strong ferromagnetic coupling (blue lines) gives much better sensitivity compared to weak coupling (orange lines) at low temperatures. (d), (e), and (f) show that the coupling strength $\mathcal{J}/\omega_0$ necessary to beat the sensitivities achievable in the weak coupling regime depends not only on $\beta$, but also on the form of the spectral density. [(a) and (d) for Ohmic spectral density $f(\Omega) = 0.001 \times \Omega$; (b) and (e) for white noise $f(\Omega) = 0.001$; (c) and (f) for 1/f-noise $f(\Omega) =  0.001/\Omega$.] \label{fig:FI_vs_beta_J_sd}}
\end{center}
\end{figure}

\subsubsection{Sensitivity vs. inverse temperature}

For weak coupling and strong anti-ferromagnetic coupling, the sensitivity $\max_t S$ saturates at a finite value as the temperature decreases ($\beta$ increases), as shown in the green and orange lines of Figs. \ref{fig:FI_vs_beta_J_sd}(a), \ref{fig:FI_vs_beta_J_sd}(b) and \ref{fig:FI_vs_beta_J_sd}(c). In contrast, for strong ferromagnetic coupling the sensitivity does not saturate, but keeps increasing as temperature decreases. From Eq. \ref{eq:max_sens} we can calculate the value at which the sensitivity saturates in the low-temperature limit: $\max_t S \stackrel{\beta\to\infty}{\longrightarrow} N / (e\gamma^-)$ in the weak coupling regime and $\max_t S \stackrel{\beta\to\infty}{\longrightarrow} N / (e\gamma^- + e\gamma^+)$ in the strong anti-ferromagnetic coupling regime. For strong ferromagnetic coupling, however, the low-temperature approximation of Eq. \ref{eq:max_sens} gives: \begin{equation} \max_t S \stackrel{\beta \gg 1}{\approx} \frac{N}{e [\gamma^- e^{-\hbar\beta|\mathcal{J}-\omega|} + \gamma^+ e^{-\hbar\beta|\mathcal{J}+\omega|}]} , \label{eq:high_beta_approx} \end{equation} which shows that (for large $\beta$) the sensitivity increases exponentially with increasing $\beta$. In the zero-temperature limit of the strong ferromagnetic coupling regime, the average decay rate vanishes $\Gamma \stackrel{\beta\to\infty}{\longrightarrow} 0$ (since $\bar{n}^{\pm} \stackrel{\beta\to\infty}{\longrightarrow} 0$) so that the sensitivity $S \stackrel{\beta\to\infty}{\longrightarrow} N \mathcal{N} t$ increases linearly with the sensing time $t$. For example, if we have a single cluster with $\mathcal{N} = N$ spins initially prepared in the $N$-spin maximally entangled state we achieve the Heisenberg limit $S_\text{HL} = N^2 t$, despite the interaction with the environment.

\subsubsection{Sensitivity vs. collective coupling strength}

The approximation in Eq. \ref{eq:high_beta_approx} is valid in the low temperature limit of the strong ferromgnetic coupling regime, but more generally it is valid when $\mathcal{J} \pm \omega  \gg 1/\hbar\beta$. This indicates that for sufficiently large $\mathcal{J}$, the sensitivity is well approximated by Eq. \ref{eq:high_beta_approx} and increases exponentially with $\mathcal{J}$. This is shown in the $\mathcal{J}\gg\omega$ stong ferromagnetic coupling region of Figs. \ref{fig:FI_vs_beta_J_sd}(d), \ref{fig:FI_vs_beta_J_sd}(e) and \ref{fig:FI_vs_beta_J_sd}(f), for three different choices of spectral density function $f(\Omega)$.

In some practical settings, the $\mathcal{J} \gg \omega$ regime may be inaccessible. An interesting question then is: how strong does the collective coupling $\mathcal{J}$ have to be to give an advantage in sensitivity over, say, a non-interacting ($\mathcal{J} = 0$) probe spin system. Comparing Figs. \ref{fig:FI_vs_beta_J_sd}(d), \ref{fig:FI_vs_beta_J_sd}(e) and \ref{fig:FI_vs_beta_J_sd}(f) shows that the answer to this question is strongly dependent on the inverse temperature $\beta$ and on the form of the spectral density function $f(\Omega)$. For an Ohmic spectral density function, Fig. \ref{fig:FI_vs_beta_J_sd}(d) shows that increasing the collective coupling $\mathcal{J}$ between spins always leads to an improved sensitivity. For white noise or for 1/f-noise, on the other hand, Figs. \ref{fig:FI_vs_beta_J_sd}(e) and \ref{fig:FI_vs_beta_J_sd}(f) show that interactions between the spins give improved sensitivity (compared to the non-interacting case, for example) only if the collective coupling $\mathcal{J}$ is larger than some critical value that depends on the inverse temperature $\beta$.

This dependence of the sensitivity on the form of the spectral density function can be partially understood by calculating the sensitivity in the region $\mathcal{J} \approx \pm \omega$. For example: \begin{equation}  \lim_{\mathcal{J}\searrow\omega} \Gamma = 2\pi \lim_{\mathcal{J}\searrow\omega} \frac{f(|\mathcal{J}-\omega|)}{\hbar\beta |\mathcal{J}-\omega|} + \frac{2\pi f(2\mathcal{J})}{e^{\hbar\beta 2\mathcal{J}} - 1} . \label{eq:Gamma_J_near_w} \end{equation} As $\mathcal{J}$ approaches $\omega$ from above, the first term in Eq. \ref{eq:Gamma_J_near_w} diverges if the spectral density function is sub-Ohmic [i.e. if $f(\Omega) \propto \Omega^k$ for $k < 1$] but is finite if the spectral density function is Ohmic or super-Ohmic [i.e. if $f(\Omega) \propto \Omega^k$ for $k \geq 1$]. Since the sensitivity is inversely proportional to $\Gamma$, this explains the sharp decrease to zero sensitivity around $\mathcal{J}\approx\omega$ for the sub-Ohmic spectral density functions in Figs. \ref{fig:FI_vs_beta_J_sd}(e) and \ref{fig:FI_vs_beta_J_sd}(f).

\subsubsection{Sensitivity vs. cluster size}

The collective coupling $\mathcal{J}$ depends on the cluster size $\mathcal{N}$. This implies that the sensitivity also depends implicitly on $\mathcal{N}$. In practice, a challenging aspect of the sensing protocol is the preparation and readout of the $\mathcal{N}$-spin entangled states, especially if the cluster size $\mathcal{N}$ is large. It is thus interesting to ask how changes in $\mathcal{N}$ affects the sensitivity.

\begin{figure}
\begin{center}
\includegraphics[width=\columnwidth]{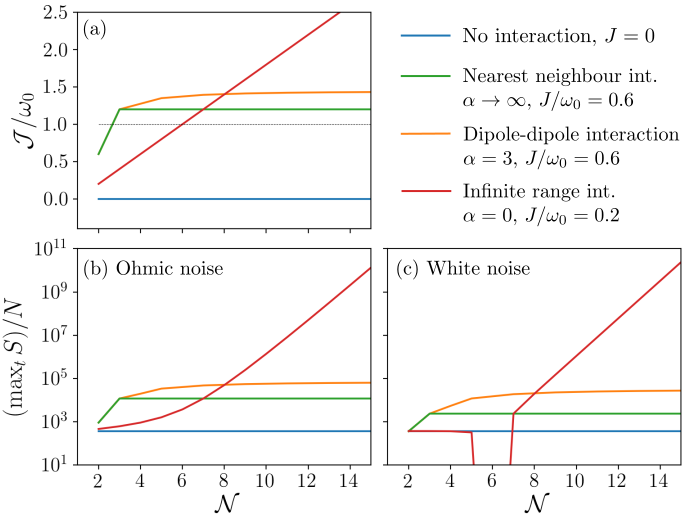} 
\caption{(a) For short-range interactions (blue, green and orange lines), the collective coupling strength $\mathcal{J}/\omega_0$ is relatively constant for $\mathcal{N}>3$. (b) and (c) If the coupling is short-range, the sensitivity is already close to its maximum value for small cluster size. If the coupling is long-range, however, sensitivity can increase with $\mathcal{N}$. [(b) is for Ohmic spectral density $f(\Omega) = 0.001 \times \Omega$; (c) is for white noise $f(\Omega) = 0.001$. Both assume $\beta = 10/\hbar\omega_0$.] \label{fig:FI_vs_beta}}
\end{center}
\end{figure}

In Fig. \ref{fig:FI_vs_beta}(a) we plot the collective coupling strength $\mathcal{J}$ as a function of the cluster size $\mathcal{N}$ for several examples. We can see that for short-range interactions [the green ($\alpha \to \infty$) and orange ($\alpha = 3$) lines], the collective coupling strength does not increase significantly as $\mathcal{N}$ increases beyond $\mathcal{N} = 3$. This is because for short-range interactions the dominant contribution to a spin's collective coupling is its coupling to its two nearest neighbours. In contrast, if the interactions are long-range, distant spins will also have a significant contibution to a spin's collective coupling, so that the collective coupling strength increases with increasing cluster size, as shown for infinite range coupling [the red line ($\alpha = 0$)] in Fig. \ref{fig:FI_vs_beta}(a). 

Since for short-range coupling the collective coupling changes relatively little for $\mathcal{N} > 3$, a large cluster size $\mathcal{N}$ (corresponding to preparation of a large maximally entangled state) does not give a substantial advantage in sensitivity compared to more clusters of smaller size $\mathcal{N} = 3$ [as illustrated in the green and orange lines, Fig. \ref{fig:FI_vs_beta}(b)]. This has important experimental implications since smaller entangled states are typically easier to prepare than large entangled states. The optimal sensing time $t_\text{opt}$ (Eq. \ref{eq:max_sens}), however, does depend on $\mathcal{N}$ and is longer for a smaller cluster size. 

If the coupling between spins is long-range, however, the collective coupling strength can increase as the cluster size increases [see the red line, Fig. \ref{fig:FI_vs_beta}(a)], resulting in an improved senstivity for a larger value of $\mathcal{N}$ [see the red line, Fig. \ref{fig:FI_vs_beta}(b)]. In the example of infinite-range coupling ($\alpha = 0$), if $\mathcal{N}$ is large enough we can approximate $\mathcal{J}\pm\omega = (\mathcal{N} - 1)J \pm \omega \approx \mathcal{N}J$ so that $\bar{n}^{\pm} \approx \exp (-\hbar\beta \mathcal{N} J)$. This means that if we are in the strong ferromagnetic couplng regime, the sensitivity $\max_{t} S \sim N\exp\left( \hbar\beta \mathcal{N} J \right)$ and the optimal sensing time $t_\text{opt} \sim \exp\left( \hbar\beta \mathcal{N} J \right) / \mathcal{N}$ increase exponentially in the cluster size $\mathcal{N}$. When $\mathcal{N} = \mathcal{O}(N)$, this raises an interesting point about the use of the phrase ``Heisenberg scaling'' in quantum metrology: Since the Heisenberg limit is $S_\text{HL} = N^2 t$ the scaling $S \propto N^2$ is often referred to as ``Heisenberg scaling''; however, in principle, $\max_{t } S$ can grow faster than $N^2$ if the optimal sensing time $t_\text{opt}$ increases with the number of particles, as this example shows.

\subsection{Example: two superconducting flux qubits}

From the foregoing discussion it is clear that an experimental demonstration of enhanced sensitivity by our scheme would require $(i)$ a qubit with a coherence time that is $T_1$-limited (i.e., close to the $T_2 \leq 2T_1$ limit) and, $(ii)$ the ability to implement a strong ferromagnetic Ising coupling $\mathcal{J}_i > \omega$ with other qubits. A minimal experimental demonstration could be achieved with a two-qubit system that satisfies these two conditions. As a candidate system, we consider two superconducting flux qubits. It has been demonstrated in several recent experiments \cite{Yos-06, Byl-11, Yan-16} that the first requirement can be met with such qubits, through the use of dynamical decoupling. The second condition can also be satisfied, since a strong ferromagnetic interaction between flux qubits has also been demonstrated experimentally \cite{Gra-06, Har-09, Lan-14}. Although both requirements have not, as yet, been implemented in a single experiment, it may be possible with future advances in the engineering of superconducting systems. In this section we choose parameters from the experiments cited above to estimate the potential gain in sensitivity with the scheme outlined in this paper.

The experiments in Refs. \cite{Byl-11, Yan-16} employ a Carr-Purcell-Meiboom-Gill (CPMG) pulse sequence in order to extend the qubit coherence time to its $T_2 \leq 2T_1$ limit. This consists of $\pi$-pulses around the $x$-axis of each qubit at the times $t_j = j t_\text{pulse}$, where $t_\text{pulse}$ is the interpulse duration and $j = 0,1,...,m$. A sensing experiment under these conditions cannot be used to precisely estimate a static parameter $\omega=\omega_0 + \Delta\omega$, since the $\pi$-pulse at $t=t_j$ causes the phase accumulated in the preceeding interval $[t_{j-1}, t_{j}]$ to be cancelled by the phase accumulated in the following interval $[t_j, t_{j+1}]$. However, if the parameter of interest is oscillating at the same frequency as the pulses are applied, the accumulated phase in each interval $[t_j,t_{j+1}]$ has the same sign and the parameter can be estimated with high sensitivity \cite{Deg-17}. Therefore, when dynamical decoupling is employed we should replace $\omega$ in our Hamiltonian Eq. \ref{eq:H_spins} with the time-dependent parameter $\omega (t) = \alpha(t)\left[ \omega_0 + \Delta\omega \sin(2\pi t/t_\text{pulse}) \right]$. Here, $\alpha(t)$ is a result of the $\pi$-pulses and takes the values $+1$ ($-1$) if the time $t$ is in the interval $[t_j, t_{j+1}]$ with $j$ even (odd). Crucially, the $\pi$-pulses do not alter the qubit-qubit interaction term, since $\left(\hat{\sigma}^x \otimes \hat{\sigma}^x \right)\left(\hat{\sigma}^z \otimes \hat{\sigma}^z \right)\left(\hat{\sigma}^x \otimes \hat{\sigma}^x \right) = \hat{\sigma}^z \otimes \hat{\sigma}^z$, so that the robustness in the presence of strong ferromagnetic coupling is maintained.

With the time dependent $\omega (t)$, the derivation of the sensitivity is similar to the time-independent case, with the final sensitivity decreased by a factor of $(2/\pi)^2$ due to the fact that the signal oscillates rather than being maintained at its maximum value $\Delta\omega$ \cite{Deg-17}.

Recent experimental results indicate that the spectral density is dominated by 1/f-noise at low qubit frequencies, but that Ohmic, and other types of noise become significant at larger qubit frequencies \cite{Yan-16}. This results in a $T_1$ time that depends on the qubit frequency. From the experimental values, we estimate $\gamma^{+} = \gamma^{-} = 1/T_1 \approx 1 / (30 \, \mu\text{s}) $ when $\omega_0 = 5 \text{ GHz}$ and $J = 0$. Since these parameters are in the weak coupling regime we can estimate the optimised sensitivity in this case as: \begin{equation} \max_t S = \frac{(2/\pi)^2N}{e\gamma (2\bar{n} + 1)} \approx 7 \times 10^{-6} \text{ Hz}^{-1} , \label{eq:S_max_WC_2} \end{equation} where we have assumed a temperature of $T_\text{env} = 20 \text{ mK}$.

If, however, the qubits are both at the frequency $\omega_0 = 2 \text{ GHz}$ and are coupled at $J = 5 \text{ GHz}$, the experimental data suggests that we can use the values $\gamma^{-} = 1/T_1 \approx 1/(20 \, \mu\text{s})$ when the qubit frequency is $|J - \omega| = 3 \text{ GHz}$, and $\gamma^{+} = 1/T_1 \approx 1/(20 \, \mu\text{s})$ when the qubit frequency is $|J+\omega| = 7 \text{ GHz}$. Since, in this case, we are in the strong ferromagnetic coupling regime, the optimised sensitivity is: \begin{equation}  \max_t S = \frac{(2/\pi)^2N}{e (\gamma^- \bar{n}^- + \gamma^+ \bar{n}^+)} \approx 11 \times 10^{-6} \text{ Hz}^{-1} , \end{equation} approximately a $50\%$ improvement in sensitivity due the strong ferromagnetic coupling between the qubits. We note that this is a minimal example of the gain that can be achieved in practice. As discussed in Sec. \ref{sec:1-d}, the gain can be increased significantly by decreasing the temperature or, more feasibly, by increasing the number of qubits that are ferromagnetically coupled. We now illustrate this by doubling the number of qubits in the example above from $N=2$ to $N=4$.

For the non-interacting case ($J_{i,j}=0$ for all $i,j$), doubling the number of qubits to $N=4$ simply doubles the optimised sensitivity to $\max_t S \approx 14 \times 10^{-6} \text{ Hz}^{-1}$. This is easily seen from the expression in Eq. \ref{eq:S_max_WC_2}, noting that when $J_{i,j}=0$ the parameters $\gamma$ and $\bar{n}$ are independent of $N$. On the other hand, if each qubit is coupled to every other qubit with $J_{i,j} = J = 5 \text{ GHz}$ then the collective coupling associated with each qubit is $\mathcal{J} = (N-1)J = 15 \text{ GHz}$. This change in the collective coupling will result in changes in the parameters $\gamma^{\pm}$ and $\bar{n}^{\pm}$. We allow for the possibility that operating a flux qubit at the high frequencies $|\mathcal{J} \pm \omega| = 15 \pm 3 \text{ GHz}$ might result in a decreased $T_1$ by choosing $\gamma^{\pm} = 1/T_1 = 1 / (2 \, \mu\text{s})$, an order of magnitude reduction of $T_1$ compared to our $N=2$ parameters. Even so, we find that the reduction in $\bar{n}^{\pm}$ for the strongly interacting qubits leads to an optimised sensitivity $\max_t S \approx 140 \times 10^{-6} \text{ Hz}^{-1}$, a factor of $10$ improvement in sensitivity compared to the non-interacting probe.

\section{Discussion}

It has been shown recently that quantum error correction can increase the robustness of frequency estimation schemes against bit-flip noise \cite{Arr-14, Kes-14, Dur-14, Und-16}. However, it appears that error correction does not significantly improve sensitivity in the presence of energy relaxation \cite{Arr-14, Mat-17}. We have shown above that robustness can be achieved by introducing strong interactions between the probes. For example, if dynamical decoupling is used to extend the probe coherence time to its fundamental limit $T_2 \leq 2T_1$, strong correlations between the probes can give a further enhancement. Other $T_1$-limited schemes, such as correlation spectroscopy \cite{Deg-17, Lar-10, Ros-17}, can also be improved by introducing interactions between the probes.











\section*{Additional Informormation}


Acknowledgements: We thank Yuichiro Matsuzaki for helpful comments.



Funding: This work was supported in part by the MEXT KAKENKHI Grant number 15H05870.

\bibliography{/Users/dooleysh/Dropbox/physics/BibTexLibrary/refs}


\end{document}